\documentclass[prl,preprint,amsmath,amssymb]{revtex4}
\usepackage{graphicx}
\usepackage[justification=centering]{caption}

\begin{document}

\title{Strong polygamy and monogamy relations for multipartite quantum systems}

\author{Zhi-Xiang Jin$^{1,2}$}
\thanks{Corresponding author: jzxjinzhixiang@126.com}
\author{Shao-Ming Fei$^{1,3}$}
\thanks{Corresponding author: feishm@cnu.edu.cn}
\affiliation{$^1$School of Mathematical Sciences, Capital Normal University,
Beijing 100048, China\\
$^2$School of Physics, University of Chinese Academy of Sciences, Beijing 100049, China\\
$^3$Max-Planck-Institute for Mathematics in the Sciences, 04103 Leipzig, Germany}

\bigskip

\begin{abstract}
Monogamy and polygamy are the most striking features of the quantum world. We investigate the monogamy and polygamy relations satisfied by all quantum correlation measures for arbitrary multipartite quantum states. By introducing residual quantum correlations, analytical polygamy inequalities are presented, which are shown to be tighter than the existing ones. Then, similar to polygamy relations, we obtain strong monogamy relations that are better than all the existing ones. Typical examples are presented for illustration.
\end{abstract}

\maketitle
\section{introduction}
Quantum correlation is one of the most important properties of quantum physics, which has been extensively studied due to its importance in quantum communication and quantum information processing. One significant property of quantum correlation is known as  monogamy.
For a tripartite system $A$, $B$ and $C$, the usual monogamy of a quantum correlation measure $Q$ implies that the correlation $Q_{A|BC}$ between $A$ and $BC$ satisfies
$Q_{A|BC}\geq Q_{AB} +Q_{AC}$. Dually, the polygamy relation is quantitatively displayed as $Q_{A|BC}\leq Q_{AB} +Q_{AC}$. It is shown that while monogamy inequalities provide an upper bound for bipartite sharability of quantum correlations in a multipartite system, the polygamy inequalities give a lower bound.
The first monogamy relation was proven for arbitrary three-qubit states based on the squared concurrence. Later, various monogamy inequalities have been established for a number of entanglement measures in multipartite quantum systems \cite{byk1,ZXN,byk2,JZX,jll,j012334,042332,gy1}. Polygamy relations are also generalized to multiqubit systems \cite{jsb} and arbitrary dimensional multipartite states \cite{JZX,jll,j012334}.

As is well known, the usual monogamy and polygamy relations are not always satisfied by any correlation measures like entanglement of formation \cite{CBHSB} quantifying the amount of entanglement required for preparation of a given bipartite quantum state. In Ref.  \cite{ZXN,JZX}, the authors have presented $\alpha$th ($\alpha\geq2$) power of concurrence and $\alpha$th ($\alpha\geq\sqrt{2}$) power of entanglement of formation satisfy the monogamy inequality for $N$-qubit. One may ask whether any measures of quantum correlations satisfy a kind of monogamy or polygamy relations.
In this paper, we first show that all quantum correlation measures satisfy some kind of polygamy relations for arbitrary multipartite quantum states.
Then we introduce the residual quantum correlations, and present tighter polygamy inequalities that are better than all the existing ones.
At last, similar to polygamy relations, we present the strong monogamy relations that are also better than the existing ones.

\smallskip
\section{strong polygamy relations for multipartite quantum systems}

Suppose $\mathcal{Q}$ is a quantum correlation measure of bipartite state. If the quantum measure $\mathcal{Q}$ satisfied the following inequality 
\begin{eqnarray}\label{}
\mathcal{Q}(\rho_{AB_1})+\mathcal{Q}(\rho_{AB_2})+\cdots+\mathcal{Q}(\rho_{AB_{N-1}})\geq \mathcal{Q}(\rho_{A|B_1B_2\cdots B_{N-1}}),
\end{eqnarray}
then we call  $\mathcal{Q}$ is polygamy for state $\rho_{AB_1B_2\cdots B_{N-1}}$, where $\rho_{AB_i}$, $i=1,...,N-1$, are the reduced states of $\rho_{AB_1B_2\cdots B_{N-1}}$. For simplicity, let $\mathcal{Q}(\rho_{AB_i})$ by $\mathcal{Q}_{AB_i}$, and $\mathcal{Q}(\rho_{A|B_1B_2 \cdots B_{N-1}})$ by $\mathcal{Q}_{A|B_1B_2\cdots B_{N-1}}$.
Similar to the monogamy relations, we define the $\mathcal{Q}$-polygamy score as follow
\begin{eqnarray}\label{}
\delta_{\mathcal{Q}}=\sum_{i=1}^{N-1}\mathcal{Q}_{AB_i}-\mathcal{Q}_{A|B_1B_2 \cdots B_{N-1}}.
\end{eqnarray}
If $\delta_{\mathcal{Q}}\geq 0$ for all states, then we say quantum measure $\mathcal{Q}$ is polygamy. For instance, it has been proved polygamous between the concurrence and concurrence of assistance for multiqubit states \cite{jsb}.

From the main result in Ref. \cite{channel}, one can find a function of a measure satisfied the polygamy relations for multipartite states. For any states $\rho_{AB_1B_2\cdots B_{N-1}}\in d\otimes d_1\otimes\cdots \otimes d_{N-1}$, exist $\beta_{\max}(\mathcal{Q})\in R$, one has 
\begin{eqnarray}\label{aq}
\mathcal{Q}^\gamma_{A|B_1B_2\cdots B_{N-1}}\leq \sum_{i=1}^{N-1}\mathcal{Q}_{AB_i}^\gamma,
\end{eqnarray}
where $0\leq \gamma\leq \beta_{\max}(\mathcal{Q})$.

Set $\beta=\beta_{\max}(\mathcal{Q})$ is the maximal value satisfied the above inequality. 
Similar to the three tangle of concurrence, for tripartite quantum states $\rho_{ABC}$, we define the residual quantum correlation as a function of $\alpha$,
\begin{eqnarray}\label{re}
\mathcal{Q}^{\alpha}_{A|B|C}=\mathcal{Q}^{\alpha}_{AB}+\mathcal{Q}^{\alpha}_{AC}-\mathcal{Q}^{\alpha}_{A|BC},~~~0\leq\alpha\leq\beta.
\end{eqnarray}
For the class of GHZ states, the equality (\ref{re}) is valid for $\beta=0$.

From the original definition in \cite{AKE,QIP}, the residual quantum correlation is defined to be $\mathcal{Q}_{A|B|C}=\mathcal{Q}_{A|BC}-\mathcal{Q}_{AB}-\mathcal{Q}_{AC}$ for some quantum correlation measures $\mathcal{Q}$ satisfying the monogamy relations $\mathcal{Q}_{A|BC}\geq \mathcal{Q}_{AB}+\mathcal{Q}_{AC}$. Generally, it is not the quantum correlation measure $\mathcal{Q}$ itself, but the
$\alpha$th power satisfies the monogamy inequality, for instance, the $\alpha$th ($\alpha\geq 2$) power of monogamy relations in \cite{ZXN}.
It is also the case for polygamy relations.
Therefore, here we use the $\alpha$th power of the quantum correlation to define the  ``residual quantum correlation".

The residual quantum correlations quantify the degree of entanglement distributions among the subsystems: the smaller of $\alpha$ in (\ref{re}), the greater degree of violation of the polygamy inequality.
Let us consider the tripartite systems. The residual quantum correlation is defined by $\mathcal{Q}^\alpha_{A|B|C}=\mathcal{Q}^\alpha_{AB}+\mathcal{Q}^\alpha_{AC}-\mathcal{Q}^\alpha_{A|BC}$ ($0\leq \alpha\leq \beta$).
For two states $\rho_{ABC}$ and $\delta_{ABC}$ such that $\mathcal{Q}^{\alpha_1}_{A|B|C}(\rho_{ABC})=\mathcal{Q}^{\alpha_2}_{A|B|C}(\delta_{ABC})=0$, $\alpha_1\leq\alpha_2$,
we have $|\mathcal{Q}(\rho_{AB})-\mathcal{Q}(\rho_{AC})|\leq |\mathcal{Q}(\delta_{AB})-\mathcal{Q}(\delta_{AC})|$.
The distribution of quantum correlation in $\rho_{ABC}$ is more averaged than that in state $\delta_{ABC}$.
For example, consider the state $|\psi\rangle=\lambda_0|000\rangle+\lambda_1e^{i{\varphi}}|100\rangle+\lambda_2|101\rangle
+\lambda_3|110\rangle+\lambda_4|111\rangle,$ where $\lambda_i\geq0,~i=0,\cdots,4$ and $\sum_{i=0}^4\lambda_i^2=1.$
We have the concurrences $C_{A|BC}=2\lambda_0\sqrt{{\lambda_2^2+\lambda_3^2+\lambda_4^2}}$, $C_{AB}=2\lambda_0\lambda_2$, and $C_{AC}=2\lambda_0\lambda_3$. Taking $\lambda_0=\lambda_1=\lambda_2=\lambda_3=\lambda_4=\frac{\sqrt{5}}{5}$, we have $\rho_{ABC}=|\psi_1\rangle\langle\psi_1|$, where $|\psi_1\rangle=\frac{\sqrt{5}}{5}|000\rangle+\frac{\sqrt{5}}{5}e^{i{\varphi}}|100\rangle+\frac{\sqrt{5}}{5}|101\rangle
+\frac{\sqrt{5}}{5}|110\rangle+\frac{\sqrt{5}}{5}|111\rangle$. One gets $C(\rho_{A|BC})^{\alpha}=(\frac{2\sqrt{3}}{5})^{\alpha}$, $C(\rho_{AB})^{\alpha}=C(\rho_{AC})^{\alpha}=(\frac{2}{5})^{\alpha}$ and $\alpha_1\thickapprox1.26185$ from $\mathcal{Q}^{\alpha_1}_{A|B|C}(\rho)=0$.
If we take $\lambda_0=\lambda_2=\frac{1}{2}$, $\lambda_1=\lambda_3=\lambda_4=\frac{\sqrt{6}}{6}$, then the state becomes $\delta_{ABC}=|\psi_2\rangle\langle\psi_2|$, where $|\psi_2\rangle=\frac{1}{2}|000\rangle+\frac{\sqrt{6}}{6}e^{i{\varphi}}|100\rangle+\frac{1}{2}|101\rangle+\frac{\sqrt{6}}{6}|110\rangle+\frac{\sqrt{6}}{6}|111\rangle$. One has $\alpha_2\thickapprox1.33770$ based on $\mathcal{Q}^{\alpha_2}_{A|B|C}(\delta_{ABC})=0$. From above, one can easily get that the entanglement distribution between the subsystems in $\rho_{ABC}$ is more averaged than that in $\delta_{ABC}$.

Consider a $d\otimes d_1\otimes d_2\otimes d_3$ state $\rho_{AB_1B_2B_3}$.
Define $\mathcal{Q}^{\alpha}_{A|B^\prime_1|B^\prime_2}=\mathrm{max}\{\mathcal{Q}^{\alpha}_{A|B_{1}|B_{2}}, \mathcal{Q}^{\alpha}_{A|B_{1}|B_{3}}, \mathcal{Q}^{\alpha}_{A|B_{2}|B_{3}}\}$,
where $B^\prime_1$ and $B^\prime_2$ stand for two of $B_1$, $B_2$ and $B_3$ such that
$\mathcal{Q}^{\alpha}_{A|B^\prime_1|B^\prime_2}=\mathrm{max}\{\mathcal{Q}^{\alpha}_{A|B_{1}|B_{2}}, \mathcal{Q}^{\alpha}_{A|B_{1}|B_{3}}, \mathcal{Q}^{\alpha}_{A|B_{2}|B_{3}}\}$.

{\bf[Theorem 1]}. For any $d\otimes d_1\otimes d_2\otimes d_3$ state $\rho_{AB_1B_2B_3}$, we have
\begin{eqnarray}\label{th2}
\mathcal{Q}^{\alpha}_{A|B_1B_2B_3}\leq \sum_{i=1}^{3}\mathcal{Q}^{\alpha}_{AB_i}-\mathcal{Q}^{\alpha}_{A|B^\prime_1|B^\prime_2},
\end{eqnarray}
for $0\leq\alpha\leq\beta$.

{\sf[Proof]}. By definition we have
\begin{eqnarray*}\label{}
 \sum_{i=1}^{3}\mathcal{Q}^{\alpha}_{AB_i}-\mathcal{Q}^{\alpha}_{A|B^\prime_{1}|B^\prime_{2}}
&&=\mathcal{Q}^{\alpha}_{AB^\prime_3}+\mathcal{Q}^{\alpha}_{A|B^\prime_1B^\prime_2}\\
&&\geq \mathcal{Q}^{\alpha}_{A|B_1B_2B_3},
\end{eqnarray*}
where $B^\prime_3$ is the complementary of $B^\prime_{1}B^\prime_{2}$ in the subsystem $B_1B_2B_3$, the equality is due to the definition of the residual quantum correlation. From (\ref{aq}) we get the inequality.
\hfill \rule{1ex}{1ex}

Concerning the parameter $\beta$ in Theorem 1, let us consider the following 4-qubit state,
\begin{eqnarray}\label{ex1}
|\psi\rangle_{AB_1B_2B_3}=&&\cos\theta_0|0000\rangle+\sin\theta_0\cos\theta_1e^{i{\varphi}}|1000\rangle+\frac{1}{2}\sin\theta_0\sin\theta_1|1010\rangle \nonumber\\&&
+\frac{3}{4}\sin\theta_0\sin\theta_1|1100\rangle+\frac{\sqrt{3}}{4}\sin\theta_0\sin\theta_1|1110\rangle,
\end{eqnarray}
where $\theta_0,\theta_1 \in[0,\frac{\pi}{2}]$. We have $C_{A|B_1B_2B_3}=2\cos\theta_0\sin\theta_0\sin\theta_1,$
$C_{AB_1}=\cos\theta_0\sin\theta_0\sin\theta_1$, $C_{AB_2}=\frac{3}{2}\cos\theta_0\sin\theta_0\sin\theta_1$ and $C_{AB_3}=C_{A|B^\prime_1|B^\prime_2}=0$.
From (\ref{th2}) we obtain $(\frac{1}{2})^\alpha+(\frac{3}{4})^\alpha\geq 1$, namely, $\alpha\leq1.507126$. Therefore, $\beta=1.507126$ is the 
largest value saturating the inequality (\ref{th2}) for the state (\ref{ex1}).

Inequality (\ref{th2}) presents a tighter polygamy relations for $0\leq\alpha\leq\beta$. Specially, inequality (\ref{th2}) is satisfied only when $\alpha=0$ for particular quantum states like the GHZ-class states.
Generalizing the conclusion of Theorem 1 to $N$ partite case, we have the following result.

{\bf[Theorem 2]}. For any $d\otimes d_1\otimes\cdots \otimes d_{N-1}$ state $\rho_{AB_1B_2\cdots B_{N-1}}$, we have
\begin{eqnarray}\label{th3}
\mathcal{Q}^{\alpha}_{A|B_1B_2\cdots B_{N-1}} \leq \sum_{i=1}^{N-1}\mathcal{Q}^{\alpha}_{AB_i}-\sum_{k=2}^{N-2}\mathcal{Q}^{\alpha}_{A|B^\prime_1|B^\prime_2|\cdots|B^\prime_k},
\end{eqnarray}
for $0\leq\alpha\leq\beta$, where $\mathcal{Q}^{\alpha}_{A|B^\prime_1|B^\prime_2|\cdots|B^\prime_{k}}=\mathrm{max}_{1\leq l\leq k+1}\{\mathcal{Q}^{\alpha}_{A|B_{1}|\cdots|\hat{B}_{l}|\cdots|B_{k+1}}\}$ (where $\hat{B}_{l}$ stands for ${B}_{l}$ being omitted in the sub-indices),
$\mathcal{Q}^{\alpha}_{A|B_{1}|B_{2}|\cdots|B_{k+1}}=\sum_{i=1}^{k+1}\mathcal{Q}^{\alpha}_{AB_i}-\mathcal{Q}^{\alpha}_{A|B_1B_2 \cdots B_{k+1}}-\sum_{i=2}^{k}\mathcal{Q}^{\alpha}_{A|B^\prime_1|B^\prime_2|\cdots|B^\prime_i}$,  $2\leq k\leq N-2$, $1\leq l\leq k+1$, $N\geq4$.

{\sf[Proof]}. We prove the theorem by induction. For $N=4$ it reduces to Theorem 1.
Suppose the Theorem 2 holds for $N=n$, i.e.,
\begin{eqnarray}\label{pfth31}
 \mathcal{Q}^{\alpha}_{A|B_1B_2\cdots B_{n-1}} \leq \sum_{i=1}^{n-1}\mathcal{Q}^{\alpha}_{AB_i}-\mathcal{Q}^{\alpha}_{A|B^\prime_{1}|B^\prime_{2}}-\cdots-\mathcal{Q}^{\alpha}_{A|B^\prime_{1}|B^\prime_{2}|\cdots|B^\prime_{n-2}}.
\end{eqnarray}
Then for $N=n+1$, we have
\begin{eqnarray*}\label{pfth32}
&& \sum_{i=1}^{n}\mathcal{Q}^{\alpha}_{AB_i}-\mathcal{Q}^{\alpha}_{A|B^\prime_{1}|B^\prime_{2}}-\cdots-\mathcal{Q}^{\alpha}_{A|B^\prime_{1}|B^\prime_{2}|\cdots|B^\prime_{n-1}}\\\nonumber
&&\geq \mathcal{Q}^{\alpha}_{A|B^\prime_{1}B^\prime_{2}\cdots B^\prime_{n-1}}+\mathcal{Q}^{\alpha}_{AB^\prime_{n}}\\\nonumber
&&\geq \mathcal{Q}^{\alpha}_{A|B_1B_2\cdots B_{n}},
\end{eqnarray*}
where $B^\prime_n$ is the complementary of $B^\prime_{1},B^\prime_{2},\cdots,B^\prime_{n-1}$ in the subsystem $B_1,B_2,\cdots,B_n$, the first inequality is due to (\ref{pfth31}). By (\ref{aq}) we get the last inequality.
\hfill \rule{1ex}{1ex}

Since the last term $\sum_{k=2}^{N-2}\mathcal{Q}^{\alpha}_{A|B^\prime_1|B^\prime_2|\cdots|B^\prime_k}$, $2\leq k\leq N-2$, $N\geq4$ in (\ref{th3}) is nonnegative,
the inequality (\ref{th3}) is always tighter than (\ref{aq}).
Let us consider the following example based on the quantum entanglement measure concurrence.
For a bipartite pure state $|\phi\rangle_{AB}$, the concurrence is $C(|\phi\rangle_{AB})=\sqrt{{2\left[1-\mathrm{Tr}(\rho_A^2)\right]}}$,
where $\rho_A$ is the reduced density matrix by tracing over the subsystem $B$, $\rho_A=\mathrm{Tr}_B(|\phi\rangle_{AB}\langle\phi|)$.
For a mixed state $\rho_{AB}=\sum_ip_i|\phi_i\rangle_{AB}\langle\phi_i|$, the concurrence is defined by the convex roof extension,
$C(\rho_{AB})=\min_{\{p_i,|\phi_i\rangle\}}\sum_ip_iC(|\phi_i\rangle)$,
where the minimum is taken over all possible decompositions of $\rho_{AB}=\sum\limits_{i}p_i|\phi_i\rangle\langle\phi_i|$, with $p_i\geq0$ and $\sum\limits_{i}p_i=1$. The concurrence of assistance is defined by $C_a(\rho_{AB})=\max_{\{p_i,|\phi_i\rangle\}}\sum_ip_iC(|\phi_i\rangle)$.
And the entanglement of assistance $\tau_a$ is given by
$\tau_a(\rho_{AB})=\sum_{m=1}^{D_1}\sum_{n=1}^{D_2}C_a((\rho_{AB})_{mn})=\sum_{m=1}^{D_1}\sum_{n=1}^{D_2}(\max\sum_ip_i|\langle\phi_i|(L_A^m\otimes L_B^n)|\phi_i^\ast\rangle|)$ \cite{022302}, where $D_1=d_1(d_1-1)/2,~D_2=d_2(d_2-1)/2$, $L_A^m=P_A^m(-|i\rangle_A\langle j|+|j\rangle_A\langle i|)P_A^m$, $L_B^n=P_B^n(-|k\rangle_B\langle l|+|l\rangle_B\langle k|)P_B^n$, and $P_A^m=|i\rangle_A\langle i|+|j\rangle_A\langle j|$, $P_B^n=|k\rangle_B\langle k|+|l\rangle_B\langle l|$ are the projections onto the subspaces spanned by $\{|i\rangle_A, |j\rangle_A\}$ and $\{|k\rangle_B, |l\rangle_B\}$, respectively. A general polygamy inequality for any multipartite pure state $|\phi\rangle_{A_1\cdots A_n}$ was established as \cite{jsb}, $\tau_a^2(|\phi\rangle_{A_1|A_2\cdots A_n})\leq \sum_{i=2}^n\tau_a^2(\rho_{A_1A_i})$,
where $\rho_{A_1A_k}$ is the reduced density matrix of subsystems $A_1A_k$ for $k=2,\cdots,n$. It has been further shown that \cite{channel},
\begin{eqnarray}\label{channel}
\tau_a^\alpha(|\phi\rangle_{A_1|A_2\cdots A_n})\leq \sum_{i=2}^n\tau_a^\alpha(\rho_{A_1A_i}),
\end{eqnarray}
where $0\leq\alpha\leq2$

{\it Example 1}. Let us consider the entanglement of assistance $\tau_a$ of the following 5-qubit pure state,
\begin{eqnarray}\label{FS}
|\psi\rangle_{AB_1B_2B_3B_4}=\frac{1}{\sqrt{5}}(|10000\rangle+|01000\rangle+|00100\rangle+|00010\rangle+|00001\rangle).
\end{eqnarray}
We have $\beta=2$, $\tau_a(|\psi\rangle_{A|B_1B_2B_3B_4})=\frac{4}{5},~\tau_a(\rho_{AB_i})=\frac{2}{5},~i=1,2,3,4$. ${\tau_a}_{A|B_i|B_j|B_k}=3(\frac{1}{2})^\alpha-(\frac{\sqrt{3}}{2})^\alpha$, $i\ne j\ne k \in\{1,2,3,4\}$. From the result (\ref{channel}) in \cite{channel}, we get $\tau_a^\alpha(|\psi\rangle_{A|B_1B_2B_3B_4}) \leq 4(\frac{2}{5})^\alpha$. From our inequality (\ref{th3}) in Theorem 2, we have $\tau_a^\alpha(|\psi\rangle_{A|B_1B_2B_3B_4})  \leq 4(\frac{2}{5})^\alpha-3(\frac{1}{2})^\alpha+(\frac{\sqrt{3}}{2})^\alpha$. Obviously, our result (\ref{th3}) is better than that in \cite{channel}, see Fig. 1.
\begin{figure}
  \centering
  \includegraphics[width=10cm]{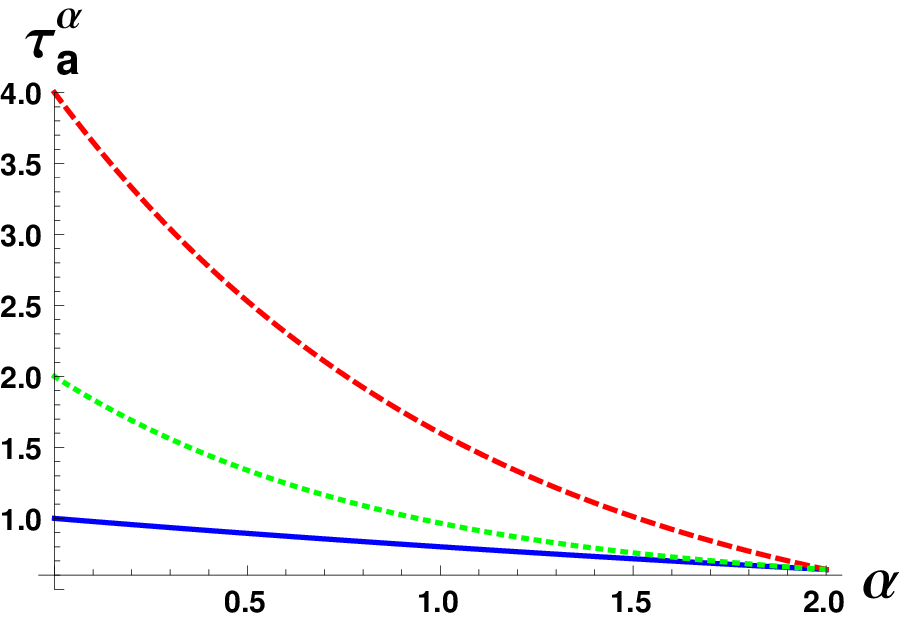}\\
  \caption{Solid (blue) line is the $\alpha$th power of $\tau_a$ under bipartition $A|B_1B_2B_3B_4$; Dashed (red) line is the upper bound in (\ref{channel}); Dotted (green) line is the upper bound in (\ref{th3}).}\label{2}
\end{figure}

In Theorems 1 and 2 we have taken into account the maximum value among $\mathcal{Q}^{\alpha}_{A|B_{1}|\cdots|\hat{B}_{l}|\cdots|B_{k}}$.
If instead of the maximum value, one just considers the mean value of $\mathcal{Q}^{\alpha}_{A|B_{1}|\cdots|\hat{B}_{l}|\cdots|B_{k}}$,
one may have the following corollary.

{\bf [Corollary 1]}. For any $d\otimes d_1\otimes\cdots \otimes d_{N-1}$ state $\rho_{A|B_1B_2\cdots B_{N-1}}$, we have
\begin{eqnarray}\label{co1}
\mathcal{Q}^{\alpha}_{A|B_1B_2\cdots B_{N-1}}\leq \sum_{i=1}^{N-1}\mathcal{Q}^{\alpha}_{AB_i}-\sum_{k=3}^{N-1}\left(\frac{1}{k}\sum_{l=1}^k\mathcal{Q}^\alpha_{A|B_{1}|\cdots|\hat{B}_{l}|\cdots|B_{k}}\right),
\end{eqnarray}
for all $0\leq\alpha\leq\beta$, $N\geq 4$, where
\begin{eqnarray}\label{col1}
 \mathcal{Q}^{\alpha}_{A|B_{1}|B_{2}|\cdots|B_{j}}=\sum_{i=1}^{j}\mathcal{Q}^{\alpha}_{AB_i}-\mathcal{Q}^{\alpha}_{A|B_1B_2\cdots B_{j}}-\sum_{k=3}^{j}\left(\frac{1}{k}\sum_{l=1}^k\mathcal{Q}^\alpha_{A|B_{1}|\cdots|\hat{B}_{l}|\cdots|B_{k}}\right),
\end{eqnarray}
$3\leq j\leq N-1$, $3\leq k\leq N-1$ and $1\leq l\leq k$.

Next, we adopt an approach used in Ref. \cite{jinzx} to improve further the above results on polygamy relations for multipartite quantum correlation measures.
First, we give a Lemma.

{\bf [Lemma 1]}. For any $d_1\otimes d_2\otimes d_3$ mixed state $\rho_{ABC}$, if $\mathcal{Q}_{AB}\geq \mathcal{Q}_{AC}$, we have
\begin{equation}\label{la1}
  \mathcal{Q}^\alpha_{A|BC}\leq  \mathcal{Q}^\alpha_{AB}+L\mathcal{Q}^\alpha_{AC},
\end{equation}
for all $0\leq\alpha\leq\beta$, where $L=(2^\frac{\alpha}{\beta}-1)$.

{\sf[Proof]}. For arbitrary $d_1\otimes d_2\otimes d_3$ tripartite state $\rho_{ABC}$.
If $\mathcal{Q}_{AB}\geq \mathcal{Q}_{AC}$, we have
\begin{eqnarray*}
  \mathcal{Q}^\alpha_{A|BC}&&\leq (\mathcal{Q}^\beta_{AB}+\mathcal{Q}^\beta_{AC})^{\frac{\alpha}{\beta}}=\mathcal{Q}^\alpha_{AB}\left(1+\frac{\mathcal{Q}^\beta_{AC}}{\mathcal{Q}^\beta_{AB}}\right)^{\frac{\alpha}{\beta}} \\
   && \leq \mathcal{Q}^\alpha_{AB}\left[1+(2^\frac{\alpha}{\beta}-1)\left(\frac{\mathcal{Q}^\beta_{AC}}{\mathcal{Q}^\beta_{AB}}\right)^{\frac{\alpha}{\beta}}\right]\\
   && =\mathcal{Q}^\alpha_{AB}+(2^\frac{\alpha}{\beta}-1)\mathcal{Q}^\alpha_{AC},
\end{eqnarray*}
where the first inequality is due to (\ref{aq}), the second inequality is due to the inequality $(1+t)^x\leq 1+(2^x-1)t^x$ for $0\leq x\leq1,~0\leq t\leq1$.
\hfill \rule{1ex}{1ex}

In the above Lemma, without loss of generality, we have assumed that $\mathcal{Q}_{AB}\geq \mathcal{Q}_{AC}$, as the subsystems
$A$ and $B$ are equivalent. Moreover, in the proof of the Lemma 1 we have assumed $\mathcal{Q}_{AB}>0$.
If $\mathcal{Q}_{AB}=0$ and $\mathcal{Q}_{AB}\geq \mathcal{Q}_{AC}$, then $\mathcal{Q}_{AB}=\mathcal{Q}_{AC}=0$. The upper bound is trivially zero.
Generalizing the Lemma 1 to multipartite quantum systems, we have the following Theorem.

{\bf[Theorem 3]}. For any $d\otimes d_1\otimes\cdots \otimes d_{N-1}$ state $\rho_{AB_1\cdots B_{N-1}}$, if
${\mathcal{Q}_{AB_i}}\geq {\mathcal{Q}_{A|B_{i+1}\cdots B_{N-1}}}$ for $i=1, 2, \cdots, m$, and
${\mathcal{Q}_{AB_j}}\leq {\mathcal{Q}_{A|B_{j+1}\cdots B_{N-1}}}$ for $j=m+1,\cdots,N-2$,
$\forall$ $1\leq m\leq N-3$, $N\geq 4$, we have
\begin{eqnarray}\label{th4}
\mathcal{Q}^\alpha_{A|B_1B_2\cdots B_{N-1}}\leq &&\mathcal{Q}^\alpha_{AB_1}
+L \mathcal{Q}^\alpha_{AB_2}+\cdots+L^{m-1}\mathcal{Q}^\alpha_{AB_m}\\\nonumber
&&+L^{m+1}(\mathcal{Q}^\alpha_{AB_{m+1}}
 +\cdots+\mathcal{Q}^\alpha_{AB_{N-2}})
+L^{m}\mathcal{Q}^\alpha_{AB_{N-1}},
\end{eqnarray}
for all $0\leq\alpha\leq\beta$, where $L=(2^\frac{\alpha}{\beta}-1)$.

{\sf[Proof]}. By using the Lemma 1 repeatedly, one gets
\begin{eqnarray}\label{pfth41}
 \mathcal{Q}^{\alpha}_{A|B_1B_2\cdots B_{N-1}}&&\leq \mathcal{Q}^{\alpha}_{AB_1}+L\mathcal{Q}^{\alpha}_{A|B_2\cdots B_{N-1}}\\\nonumber
&&\leq \mathcal{Q}^{\alpha}_{AB_1}+L\mathcal{Q}^{\alpha}_{AB_2}
 +L^2\mathcal{Q}^{\alpha}_{A|B_3\cdots B_{N-1}}\\ \nonumber
 &&\leq\cdots\leq \mathcal{Q}^{\alpha}_{AB_1}+L\mathcal{Q}^{\alpha}_{AB_2}+\cdots\\\nonumber
 &&+ L^{m-1}\mathcal{Q}^{\alpha}_{AB_m}
 +L^m \mathcal{Q}^{\alpha}_{A|B_{m+1}\cdots B_{N-1}}.
\end {eqnarray}
As ${\mathcal{Q}_{AB_j}}\leq {\mathcal{Q}_{A|B_{j+1}\cdots B_{N-1}}}$ for $j=m+1,\cdots,N-2$, by (\ref{la1}) we get
\begin{eqnarray}\label{pfth42}
\mathcal{Q}^{\alpha}_{A|B_{m+1}\cdots B_{N-1}}&&\leq L\mathcal{Q}^{\alpha}_{AB_{m+1}}+\mathcal{Q}^{\alpha}_{A|B_{m+2}\cdots B_{N-1}}\nonumber\\
&&\leq L(\mathcal{Q}^{\alpha}_{AB_{m+1}}+\cdots+\mathcal{Q}^{\alpha}_{AB_{N-2})}+\mathcal{Q}^{\alpha}_{AB_{N-1}}.
\end{eqnarray}
Combining (\ref{pfth41}) and (\ref{pfth42}), we have Theorem 3.
\hfill \rule{1ex}{1ex}

Similar to the Theorem 2, (\ref{th4}) can be improved by adding a term for residual quantum correlation.
By a similar derivation to Theorem 2, we have

{\bf[Theorem 4]}. For any $d\otimes d_1\otimes\cdots \otimes d_{N-1}$ state $\rho_{AB_1\cdots B_{N-1}}$, if
${\mathcal{Q}_{AB_i}}\geq {\mathcal{Q}_{A|B_{i+1}\cdots B_{N-1}}}$ for $i=1, 2, \cdots, m$, and
${\mathcal{Q}_{AB_j}}\leq {\mathcal{Q}_{A|B_{j+1}\cdots B_{N-1}}}$ for $j=m+1,\cdots,N-2$,
$\forall$ $1\leq m\leq N-3$, $N\geq 4$, we have
\begin{eqnarray}\label{}\label{th5}
\mathcal{Q}^\alpha_{A|B_1B_2\cdots B_{N-1}}\leq \sum_{i=1}^{N-1}\mathcal{\hat{Q}}^{\alpha}_{AB_i}-\sum_{k=2}^{N-2}\mathcal{\hat{Q}}^{\alpha}_{A|B^\prime_1|B^\prime_2|\cdots|B^\prime_{k}},
\end{eqnarray}
for all $0\leq\alpha\leq\beta$, where
$\mathcal{\hat{Q}}^\alpha_{AB_1}=\mathcal{Q}^\alpha_{AB_1}$, $\mathcal{\hat{Q}}^\alpha_{AB_2}=L\mathcal{Q}^\alpha_{AB_2}$, $\cdots$, $\mathcal{\hat{Q}}^\alpha_{AB_m}=L^{m-1}\mathcal{Q}^\alpha_{AB_m}$, $\mathcal{\hat{Q}}^\alpha_{AB_{m+1}}=L^{m+1}\mathcal{Q}^\alpha_{AB_{m+1}}$, $\cdots$, $\mathcal{\hat{Q}}^\alpha_{AB_{N-2}}=L^{m+1}\mathcal{Q}^\alpha_{AB_{N-2}}$, $\mathcal{\hat{Q}}^\alpha_{AB_{N-1}}=L^{m}\mathcal{Q}^\alpha_{AB_{N-1}}$, $L=(2^\frac{\alpha}{\beta}-1)$.
The residual quantum correlation term $\mathcal{\hat{Q}}^{\alpha}_{A|B^\prime_1|B^\prime_2|\cdots|B^\prime_{k-1}}=\mathrm{max}_{1\leq l\leq k}\{\mathcal{\hat{Q}}_{A|B_{1}|\cdots|\hat{B}_{l}|\cdots|B_{k}}\}$,
$\mathcal{\hat{Q}}^{\alpha}_{A|B_{1}|B_{2}|\cdots|B_{k}}=\sum_{i=1}^{k}\mathcal{\hat{Q}}^{\alpha}_{AB_i}-\mathcal{Q}^{\alpha}_{A|B_1B_2\cdots B_{k}}-\sum_{i=2}^{k-1}\mathcal{\hat{Q}}^{\alpha}_{A|B^\prime_1|B^\prime_2|\cdots|B^\prime_i}$, $2\leq k\leq N-2$, $1\leq l\leq k$.

As an example, let us consider consider again the the concurrence of the state (\ref{FS}).
From our inequality (\ref{th3}) in Theorem 2, we have $\tau_a^\alpha(|\psi\rangle_{A|B_1B_2B_3B_4})  \leq 4(\frac{2}{5})^\alpha-3(\frac{1}{2})^\alpha+(\frac{\sqrt{3}}{2})^\alpha$. From the inequality (\ref{th5}) in Theorem 4, we have $\tau_a^\alpha(|\psi\rangle_{A|B_1B_2B_3B_4})  \leq 3(\frac{2\sqrt{2}}{5})^\alpha-2(\frac{2}{5})^\alpha-2(\frac{1}{2})^\frac{\alpha}{2}+(\frac{1}{2})^\alpha+(\frac{\sqrt{3}}{2})^\alpha$. Obviously, the inequality (\ref{th5}) is better than the inequality in \cite{channel}. We see in Fig. 2 that the bound (\ref{th3}) is improved.

\begin{figure}
  \centering
  \includegraphics[width=10cm]{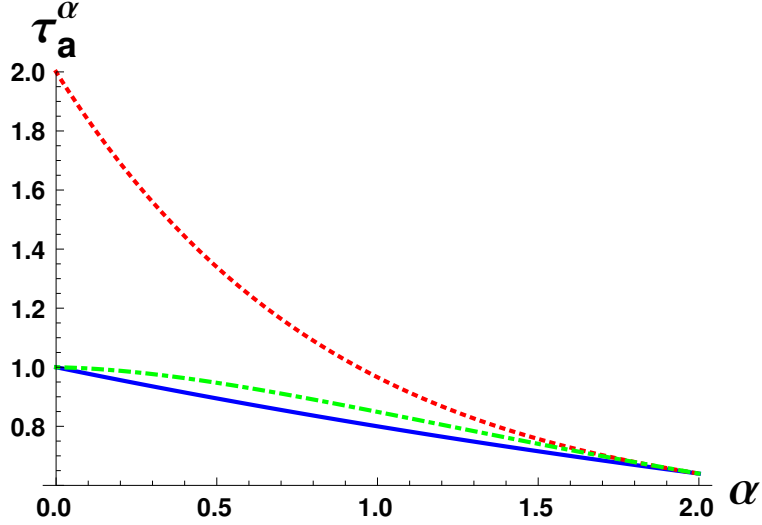}\\
  \caption{Solid (blue) line is the $\alpha$th power of $\tau_a$ under bipartition $A|B_1B_2B_3B_4$; Dashed (red) line is the upper bound (\ref{th3}); Dotted (green) line is the upper bound in (\ref{th5}).}\label{2}
\end{figure}

\smallskip

\section{strong monogamy relations for multipartite quantum systems}

We now study the monogamy relations for multipartite states. The monogamy relations limit the distributions of quantum correlations among the multipartite systems and play an important role
in secure quantum cryptography \cite{MP} and in condensed matter physics such as the $n$-representability problem for fermions \cite{AV}.
Monogamy relations of entanglement for multiqubit some higher-dimensional quantum systems have been investigated in terms of various entanglement measures \cite{ZXN,JZX,AKE,SSS,jll}.
Some of the quantum measures, however, do not satisfy the monogamy inequality, even for the pure three qubit states \cite{GLGP, RPAK}.
In \cite{SPAU} the authors give a monogamy power, $x_{\min}(\mathcal{Q})\in R$, for arbitrary dimensional tripartite states, $\mathcal{Q}$ satisfies
\begin{eqnarray}\label{aqm}
\mathcal{Q}^y_{A|BC}\geq\mathcal{Q}^y_{AB}+\mathcal{Q}^y_{AC},
\end{eqnarray}
here $y\geq x_{\min}(\mathcal{Q})$.

In the following, denoting $x=x_{\min}(\mathcal{Q})$ is the minimal value satisfied (\ref{aqm}).
Inequality (\ref{aqm}) has been generalized to the $N$ partite case for all measures of quantum correlations \cite{jzxoc},
\begin{eqnarray}\label{monogamy1}
\mathcal{Q}^y_{A|B_1B_2\cdots B_{N-1}}\geq \sum_{i=1}^{N-1}\mathcal{Q}_{AB_i}^y,
\end{eqnarray}
for $y\geq x$, $N\geq 3$. (\ref{monogamy1}) has been further improved such that for $y\geq x$, if
${\mathcal{Q}_{AB_i}}\geq {\mathcal{Q}_{A|B_{i+1}\cdots B_{N-1}}}$ for $i=1, 2, \cdots, m$, and
${\mathcal{Q}_{AB_j}}\leq {\mathcal{Q}_{A|B_{j+1}\cdots B_{N-1}}}$ for $j=m+1,\cdots,N-2$,
$\forall$ $1\leq m\leq N-3$, $N\geq 4$, then \cite{jzxoc},
\begin{eqnarray}\label{monogamy2}
\mathcal{Q}^y_{A|B_1B_2\cdots B_{N-1}}\geq \sum_{i=1}^{N-1}\mathcal{\hat{Q}}^y_{AB_i}+\sum_{k=2}^{N-2}\mathcal{\hat{Q}}^y_{A|B^\prime_1|B^\prime_2|\cdots|B^\prime_{k}},
\end{eqnarray}
for all $y\geq x$, $\mathcal{\hat{Q}}^y_{AB_1}=\mathcal{Q}^y_{AB_1}$, $\mathcal{\hat{Q}}^y_{AB_2}=K \mathcal{Q}^y_{AB_2}$, $\cdots$, $\mathcal{\hat{Q}}^y_{AB_m}=K^{m-1}\mathcal{Q}^y_{AB_m}$, $\mathcal{\hat{Q}}^y_{AB_{m+1}}=K^{m+1}\mathcal{Q}^y_{AB_{m+1}}$, $\cdots$, $\mathcal{\hat{Q}}^y_{AB_{N-2}}=K^{m+1}\mathcal{Q}^y_{AB_{N-2}}$, $\mathcal{\hat{Q}}^y_{AB_{N-1}}=K^{m}\mathcal{Q}^y_{AB_{N-1}}$ and $K=\frac{y}{x}$.
The residual quantum correlation term $\mathcal{\hat{Q}}^y_{A|B^\prime_1|B^\prime_2|\cdots|B^\prime_{k-1}}=\mathrm{max}_{1\leq l\leq k}\{\mathcal{\hat{Q}}_{A|B_{1}|\cdots|\hat{B}_{l}|\cdots|B_{k}}\}$ (where $\hat{B}_{l}$ stands for ${B}_{l}$ being omitted in the sub-indices),
$\mathcal{\hat{Q}}^y_{A|B_{1}|B_{2}|\cdots|B_{k}}=\mathcal{Q}^y_{A|B_1B_2\cdots B_{k}} -\sum_{i=1}^{k}\mathcal{\hat{Q}}^y_{AB_i}-\sum_{i=2}^{k-1}\mathcal{\hat{Q}}^y_{A|B^\prime_1|B^\prime_2|\cdots|B^\prime_i}$, $2\leq k\leq N-2$, $1\leq l\leq k$.

In fact, as a kind of characterization of the quantum correlation distribution among the subsystems, the monogamy inequalities satisfied by the quantum correlations can be further refined and become tighter.

{\bf [Lemma 2]}. For any $d_1\otimes d_2\otimes d_3$ mixed state $\rho_{ABC}$, if $\mathcal{Q}_{AB}\geq \mathcal{Q}_{AC}$, we have
\begin{equation}\label{la}
  \mathcal{Q}^y_{A|BC}\geq  \mathcal{Q}^y_{AB}+L\mathcal{Q}^y_{AC},
\end{equation}
for all $y\geq x$, where $L=(2^\frac{y}{x}-1)$.

{\sf[Proof]}. For arbitrary $d_1\otimes d_2\otimes d_3$ tripartite state $\rho_{ABC}$.
If $\mathcal{Q}_{AB}\geq \mathcal{Q}_{AC}$, we have
\begin{eqnarray*}
  \mathcal{Q}^y_{A|BC}&&\geq (\mathcal{Q}^x_{AB}+\mathcal{Q}^x_{AC})^{\frac{y}{x}}=\mathcal{Q}^y_{AB}\left(1+\frac{\mathcal{Q}^x_{AC}}{\mathcal{Q}^x_{AB}}\right)^{\frac{y}{x}} \\
   && \geq \mathcal{Q}^y_{AB}\left[1+L\left(\frac{\mathcal{Q}^x_{AC}}{\mathcal{Q}^x_{AB}}\right)^\frac{y}{x}\right]\\
   && =\mathcal{Q}^y_{AB}+L\mathcal{Q}^y_{AC},
\end{eqnarray*}
where the first inequality is due to (\ref{aqm}), the second inequality is due to the inequality $(1+t)^x\geq 1+(2^x-1)t^x$ for $x\geq1,~0\leq t\leq1$ \cite{jll}.
\hfill \rule{1ex}{1ex}

{\bf[Theorem 5]}. For any $d\otimes d_1\otimes\cdots \otimes d_{N-1}$ state $\rho_{AB_1\cdots B_{N-1}}$, if
${\mathcal{Q}_{AB_i}}\geq {\mathcal{Q}_{A|B_{i+1}\cdots B_{N-1}}}$ for $i=1, 2, \cdots, m$, and
${\mathcal{Q}_{AB_j}}\leq {\mathcal{Q}_{A|B_{j+1}\cdots B_{N-1}}}$ for $j=m+1,\cdots,N-2$,
$\forall$ $1\leq m\leq N-3$, $N\geq 4$, we have
\begin{eqnarray}\label{tha}
\mathcal{Q}^y_{A|B_1B_2\cdots B_{N-1}}\geq \sum_{i=1}^{N-1}\mathcal{\widetilde{Q}}^y_{AB_i}+\sum_{k=2}^{N-2}\mathcal{\widetilde{Q}}^y_{A|B^\prime_1|B^\prime_2|\cdots|B^\prime_{k}},
\end{eqnarray}
for all $y\geq x$,
where $\mathcal{\widetilde{Q}}^y_{AB_1}=\mathcal{Q}^y_{AB_1}$, $\mathcal{\widetilde{Q}}^y_{AB_2}=L\mathcal{Q}^y_{AB_2}$, $\cdots$, $\mathcal{\widetilde{Q}}^y_{AB_m}=L^{m-1}\mathcal{Q}^y_{AB_m}$, $\mathcal{\widetilde{Q}}^y_{AB_{m+1}}=L^{m+1}\mathcal{Q}^y_{AB_{m+1}}$, $\cdots$, $\mathcal{\widetilde{Q}}^y_{AB_{N-2}}=L^{m+1}\mathcal{Q}^y_{AB_{N-2}}$, $\mathcal{\widetilde{Q}}^y_{AB_{N-1}}=L^{m}\mathcal{Q}^y_{AB_{N-1}}$,
$L=(2^\frac{y}{x}-1)$. The residual quantum correlation term $\mathcal{\widetilde{Q}}^y_{A|B^\prime_1|B^\prime_2|\cdots|B^\prime_{k-1}}=\mathrm{max}_{1\leq l\leq k}\{\mathcal{\widetilde{Q}}_{A|B_{1}|\cdots|\hat{B}_{l}|\cdots|B_{k}}\}$,
$\mathcal{\widetilde{Q}}^y_{A|B_{1}|B_{2}|\cdots|B_{k}}=\mathcal{Q}^y_{A|B_1B_2\cdots B_{k}} -\sum_{i=1}^{k}\mathcal{\widetilde{Q}}^y_{AB_i}-\sum_{i=2}^{k-1}\mathcal{\widetilde{Q}}^y_{A|B^\prime_1|B^\prime_2|\cdots|B^\prime_i}$, $2\leq k\leq N-2$, $1\leq l\leq k$.

{\sf[Proof]}. By using the Lemma 2 repeatedly, one gets
\begin{eqnarray}\label{pftha1}
 \mathcal{Q}^y_{A|B_1B_2\cdots B_{N-1}}&&\geq \mathcal{Q}^y_{AB_1}+L\mathcal{Q}^{\alpha}_{A|B_2\cdots B_{N-1}}\nonumber\\
&&\geq \mathcal{Q}^y_{AB_1}+L\mathcal{Q}^{\alpha}_{AB_2}
 +L^2\mathcal{Q}^y_{A|B_3\cdots B_{N-1}}\nonumber\\
 &&\geq\cdots\geq \mathcal{Q}^y_{AB_1}+L\mathcal{Q}^y_{AB_2}+\cdots\nonumber\\
 &&+ L^{m-1}\mathcal{Q}^y_{AB_m}
 +L^m \mathcal{Q}^y_{A|B_{m+1}\cdots B_{N-1}}.
\end {eqnarray}
As ${\mathcal{Q}_{AB_j}}\leq {\mathcal{Q}_{A|B_{j+1}\cdots B_{N-1}}}$ for $j=m+1,\cdots,N-2$, by (\ref{pfth41}) we get
\begin{eqnarray}\label{pftha2}
\mathcal{Q}^y_{A|B_{m+1}\cdots B_{N-1}}&&\geq L\mathcal{Q}^y_{AB_{m+1}}+\mathcal{Q}^y_{A|B_{m+2}\cdots B_{N-1}}\nonumber\\
&&\geq L(\mathcal{Q}^y_{AB_{m+1}}+\cdots+\mathcal{Q}^y_{AB_{N-2})}+\mathcal{Q}^y_{AB_{N-1}}.
\end{eqnarray}
Combining (\ref{pftha1}) and (\ref{pftha2}), we have
\begin{eqnarray}\label{pftha3}
\mathcal{Q}^y_{A|B_1B_2\cdots B_{N-1}}\geq \sum_{i=1}^{N-1}\mathcal{\widetilde{Q}}^y_{AB_i}.
\end{eqnarray}
Suppose that Theorem 5 holds for $N=n$, i.e.,
\begin{eqnarray}\label{pftha4}
 \mathcal{Q}^y_{A|B_1B_2\cdots B_{n-1}} \geq \sum_{i=1}^{n-1}\mathcal{\widetilde{Q}}^y_{AB_i}+\mathcal{\widetilde{Q}}^y_{A|B^\prime_{1}|B^\prime_{2}}+\cdots+\mathcal{\widetilde{Q}}^y_{A|B^\prime_{1}|B^\prime_{2}|\cdots|B^\prime_{n-2}}.
\end{eqnarray}
Then for $N=n+1$, we have
\begin{eqnarray*}
&& \sum_{i=1}^{n}\mathcal{\widetilde{Q}}^y_{AB_i}+\mathcal{\widetilde{Q}}^y_{A|B^\prime_{1}|B^\prime_{2}}+\cdots+\mathcal{\widetilde{Q}}^y_{A|B^\prime_{1}|B^\prime_{2}|\cdots|B^\prime_{n-1}}\\\nonumber
&&\leq \mathcal{\widetilde{Q}}^y_{A|B^\prime_{1}B^\prime_{2}\cdots B^\prime_{n-1}}+\mathcal{\widetilde{Q}}^y_{AB^\prime_{n}}\\\nonumber
&&\leq \mathcal{Q}^y_{A|B_1B_2\cdots B_{n}},
\end{eqnarray*}
where $B^\prime_n$ is the complementary of $B^\prime_{1}B^\prime_{2},\cdots,B^\prime_{n-1}$ in the subsystem $B_1B_2,\cdots,B_n$. The first inequality is due to (\ref{pftha4}). By (\ref{pftha3}) we get the last inequality.
\hfill \rule{1ex}{1ex}

{\it Example 2}. For the concurrence of the $W$ state,
\begin{eqnarray}\label{W}
|W\rangle_{A|B_1B_2B_3}=\frac{1}{2}(|1000\rangle+|0100\rangle+|0010\rangle+|0001\rangle),
\end{eqnarray}
we have $x=2$, $C_{AB_i}=\frac{1}{2}$, $i=1,2,3$, and $C_{A|B_1B_2}=C_{A|B_1B_3}=C_{A|B_2B_3}=\frac{\sqrt{2}}{2}$. From the inequality (\ref{monogamy2}), one has $\hat{C}^y_{A|B_1|B_2}=\hat{C}^y_{A|B_1|B_3}=\hat{C}^y_{A|B_2|B_3}=(\frac{\sqrt{2}}{2})^y-(1+\frac{y}{2})(\frac{1}{2})^y$. Hence the lower bound of $C^y_{A|B_1B_2B_3}$ is $\sum_{i=1}^3\hat{C}^y_{AB_i}+\hat{C}^y_{A|B_1|B_2}=(\frac{\sqrt{2}}{2})^y+\frac{y}{2}(\frac{1}{2})^y$. From the inequality (\ref{tha}) in Theorem 5, we have $\widetilde{C}^y_{A|B_1|B_2}=\widetilde{C}^y_{A|B_1|B_3}=\widetilde{C}^y_{A|B_2|B_3}=(\frac{\sqrt{2}}{2})^y-(\frac{1}{2})^\frac{y}{2}$. The lower bound of $C^y_{A|B_1B_2B_3}$ is $\sum_{i=1}^3\widetilde{C}^y_{AB_i}+\widetilde{C}^y_{A|B_1|B_2}=(\frac{\sqrt{2}}{2})^y+(2^\frac{y}{2}-1)(\frac{1}{2})^y$. One can see that our result is better than (\ref{monogamy2}) in \cite{jzxoc}, see Fig. 3.
\begin{figure}
  \centering
  \includegraphics[width=10cm]{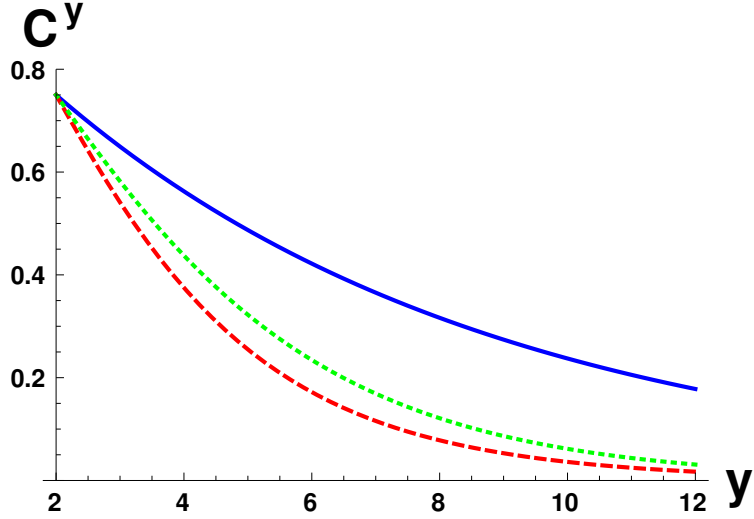}\\
  \caption{Solid (blue) line is the $y$th power of concurrence under bipartition $A$ and $B_1B_2B_3$; Dashed (red) line for the lower bound (\ref{monogamy2}) in \cite{jzxoc}; Dotted (green) line for the lower bound in (\ref{tha}).}\label{2}
\end{figure}

\section{Conclusion}
We have investigated the monogamy and polygamy relations satisfied by arbitrary quantum correlation measures for arbitrary multipartite quantum states. We have introduced the $\alpha$th $(0\leq\alpha\leq\beta)$ power of the residual quantum correlation.
In term of the residual quantum correlations, analytical polygamy inequalities have been presented, which are shown to be tighter than the existing ones.
Similarly, we have obtained the strong monogamy relations that are also better than all the existing ones. Detailed examples have been given for illustration.

\bigskip
\noindent{\bf Acknowledgments}\, \, This work is supported by the NSF of China under Grant No. 11847209, 11675113;  the Key Project of
Beijing Municipal Commission of Education (Grant No. KZ201810028042); Beijing Natural Science Foundation (Grant No. Z190005), and China Postdoctoral Science Foundation Funded Project.


\begin{thebibliography}{99}
\bibitem{byk1}Y. K. Bai, M. Y. Ye, and Z. D. Wang, Entanglement monogamy and entanglement evolution in multipartite systems. Phys. Rev. A 80, 044301(2009).
\bibitem{ZXN} X. N. Zhu and S. M. Fei, Entanglement monogamy relations of qubit systems. Phys. Rev. A 90, 024304 (2014).
\bibitem{JZX} Z. X. Jin and S. M. Fei, Tighter entanglement monogamy relations of qubit systems. Quantum Inf Process  16:77 (2017).
\bibitem{j012334} J. S. Kim, Negativity and tight constraints of multiqubit entanglement. Phys. Rev. A 97, 012334 (2018).
\bibitem{jll}  Z. X. Jin, J. Li, T. Li, S. M. Fei, Tighter monogamy relations in multiqubit systems. Phys. Rev. A 97, 032336 (2018).
\bibitem{byk2} Y. K. Bai, Y. F. Xu, and Z. D. Wang, General monogamy relation for the entanglement of formation in multiqubit systems. Phys. Rev. Lett. 113, 100503 (2014).



\bibitem{042332} J. S. Kim, Weighted polygamy inequalities of multiparty entanglement in arbitrary-dimensional quantum systems, Phys. Rev. A 97, 042332 (2018).
\bibitem{gy1} G. Gour, Y. Guo, Monogamy of entanglement without inequalities. Quantum 2, 81 (2018).



\bibitem{jsb} G. Gour, S. Bandyopadhay, and B. C. Sanders, Dual monogamy inequality for entanglement. J. Math. Phys. 48, 012108 (2007).
\bibitem{CBHSB} C. H. Bennett, H. J. Bernstein, S. Popescu, B. Schumacher, Concentrating partial entanglement by local operations. Phys. Rev. A 53, 2046 (1996).
\bibitem{channel} Z. X. Jin, S. M. Fei, Superactivation of monogamy relations for nonadditive quantum correlation measures, Phys. Rev. A 99, 032343 (2019).

\bibitem{jinzx} Z. X. Jin, S. M. Fei, Finer distribution of quantum correlations among multiqubit systems. Quantum Inf Process  18:21 (2019).

\bibitem{MP} M. Pawlowski, Security proof for cryptographic protocols based only on the monogamy of bells inequality violations. Phys. Rev. A 82, 032313 (2010).	
\bibitem{AV} A. J. Coleman and V. I. Yukalov, Reduced Density Matrices: Coulson's Challenge. Lecture Notes in Chemistry, vol. 72. Springer, Berlin (2000).

\bibitem{AKE} A. K. Ekert, Quantum cryptography based on Bell's theorem. Phys. Rev. Lett. 67 661 (1991).

\bibitem{QIP} V. Coffman, J. Kundu, W. K. Wootters, Distributed entanglement. Phys. Rev. A 61, 052306 (2000).

\bibitem{gs} Groblacher S, Jennewein T, Vaziri A, Weihs G and Zeilinger A, Experimental quantum cryptography with qutrits. New J. Phys. 8 75 (2006).

\bibitem{022302} J. S. Kim, Polygamy of entanglement in multipartite quantum systems. Phys. Rev. A 80, 022302 (2009).

\bibitem{SSS} G. Adesso, A. Serafini, F. Illuminati, Multipartite entanglement in three-mode Gaussian states of continuous-variable systems: Quantification, sharing structure, and decoherence. Phys. Rev. A 73, 032345 (2006).



\bibitem{RPAK} R. Prabhu, A. K. Pati, A. Sen(De), U. Sen, Conditions for monogamy of quantum correlations: Greenberger-Horne-Zeilinger versus W states. Phys. Rev. A 85, 040102(R) (2012).
\bibitem{GLGP} G. L. Giorgi, Monogamy properties of quantum and classical correlations. Phys. Rev. A 84, 054301 (2011) .

\bibitem{SPAU}K. Salini, R. Prabhu, A. Sen(De), and U. Sen, Monotonically increasing functions of any quantum
correlation can make all multiparty states monogamous. Ann. Phys 348, 297-305 (2014) .
\bibitem{jzxoc}Z. X. Jin, S. M. Fei, Monogamy relations of all quantum correlation measures for multipartite quantum systems. Optics Communications 446, 39-43 (2019).









\end{thebibliography}
\end{document}